\documentclass[twocolumn,showpacs,preprintnumbers,amsmath,amssymb]{revtex4}

\usepackage{graphicx}
\usepackage{dcolumn}
\usepackage{amsmath}    
\usepackage{verbatim}   
\usepackage{color}      
\usepackage{subfigure}  
\usepackage{hyperref}   
\usepackage{bm}

\begin{document}


\title{Experimental position-time entanglement with degenerate single photons}

\author{A. J. Bennett}
 \email{anthony.bennett@crl.toshiba.co.uk}
\author{D. G. Gevaux}
\author{Z. L. Yuan}
\author{A. J. Shields}%

\affiliation{%
Toshiba Research Europe Limited, Cambridge Research Laboratory,\\
208 Science Park, Milton Road, Cambridge, CB4 OGZ, U. K.}%

\author{P. Atkinson}
\author{D. A. Ritchie}
\affiliation{Cavendish Laboratory, Cambridge University,\\
Madingley Road, Cambridge, CB3 0HE, U. K.}%

\date{\today}%

\begin{abstract}
We report an experiment in which two-photon
interference occurs between degenerate single
photons that never meet. The two photons travel
in opposite directions through our fibre-optic
interferometer and interference occurs when the
photons reach two different, spatially separated,
2-by-2 couplers at the same time. We show that
this experiment is analogous to the conventional
Franson-type entanglement experiment where the
photons are entangled in position and time. We
measure wavefunction overlaps for the two photons
as high as 94 $\pm$ 3\%.
\end{abstract}


\maketitle 

\section{INTRODUCTION}

The fact that the mathematical model of quantum
mechanics suggests an object can exist in a
superposition state, for instance in two
different places at the same time, is one of the
most surprising realisations people have when
first learning quantum mechanics. The fact that a
number of different particles can exist in a
superposition-state is the logical extension made
by Einstein, Podolski and Rosen (EPR) \cite{EPR}
and then Greenberger, Horne and Zeilinger
\cite{GHZ}. Often the EPR experiment is explained
to students as if the two particles are
"entangled" at the point of emission, usually in
the polarization-basis. However, it can be shown
that two separate single particles which are
certainly not entangled at the point of emission,
but are identical, can also display entanglement
if an appropriate experiment is designed which
post-selects events where they cannot be
distinguished. Here we interfere two single
photons from a single-photon-source using only a
Mach-Zehnder interferometer (MZI), in which the
photons actually travel in opposite directions
and interfere on separate couplers, at different
times.

\begin{figure}
\includegraphics[width = 80mm]{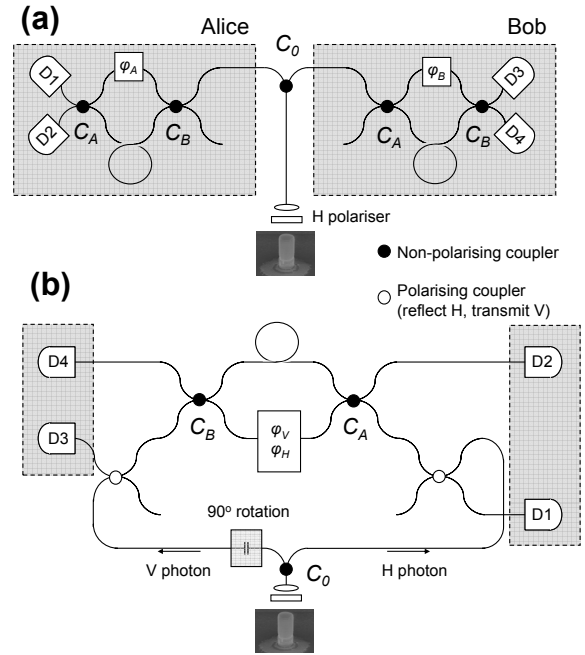}
\caption{\label{fig:epsart} (a) Conventional
layout of Franson-type experiment to demonstrate
entanglement in position and time between two
identical particles emitted by the source. (b)
our experimental apparatus. The component in the
short arm of the Mach-Zehnder interferometer
induces a polarization dependent phase shift
$\phi_{V}$ for V (leftward) traveling photons and
$\phi_{H}$ for H (rightward) traveling photons.}
\label{Fig1}
\end{figure}

\section{EXPERIMENTAL ARRANGEMENT}

The experiment, similar to that proposed by
Franson \cite{Franson}, is shown schematically in
Fig.1(a). The experiment is set up so that the
two indistinguishable photons emitted by the
source have a time separation $\Delta t$ which is
equal to the difference in the delays induced by
each arm in each MZI, $n(l_{A}-s_{A})/c$, where
$n$ is the refractive index, $l_{A}$ and $s_{A}$
the lengths of the long and short arms in Alice's
MZI. $\Delta t$ is an order of magnitude longer
than the coherence time of the photons and thus
no second-order interference can occur when the
paths are recombined \cite{Brendel, Kwiat}. The
only way a coincidence can occur is if the first
photon takes the long path and is delayed by
$\Delta t$ relative to the second photon, taking
the short path. In this case, it will not be
possible for two detectors in Alice's
interferometer (D1 and D2) to distinguish the
events where both photons arrive at the detectors
at the same time: either both were reflected at
coupler $C_{A}$, or both were transmitted. If the
coupler $C_{A}$ has equal reflection and
transmission coefficients fourth-order
interference ensures these two events cancel out
and no coincident detection events for D1 and D2.
This two photon interference effect requires only
that $\Delta t$ is approximately equal to
$n(l_{A}-s_{A})/c$ to within the length of the
photon wavepacket. In other words, the
observation of fourth order interference does not
require a fixed phase relationship between each
pair of photons or wavelength-scale stability of
$(l_{A}-s_{A})$, and so is independent of the
phase $\phi_{A}$. A similar requirement is needed
to observe two-photon interference at Bob's end
of the apparatus. Quantum-dot single-photon
sources that can lead to this two-photon
interference effect have now been reported by
several groups \cite{SantoriNature,
BennettOptExp2, Varoutsis}. Already a number of
experiments have used this effect to implement
simple quantum protocols such as teleportation
\cite{Fattal1} and polarization entanglement
generation \cite{Fattal2}. It is sometimes argued
that this interference effect, widely studied
with photons, is due to their bosonic nature and
results from the photons "sticking together" when
they collide at the coupler $C_{A}$. However,
there is no interaction between photons. The
Franson entanglement experiment shows that in
actual fact the photons can interfere even if
they are incident on separate couplers, such that
they are never in the same place at the same
time. Considering detectors $\{i,j\}$ it can be
shown that the probability of a coincidence,
$\Gamma\{i,j\}$ is given by:

\begin{subequations}
\begin{align}
\label{equation1} \Gamma\{1,3\} \propto
R_{A}T_{A}R_{B}T_{B}[2 + 2 \gamma^{2}cos(\phi_{A} -\phi_{B})]\\
\Gamma\{1,4\} \propto R_{A}T_{A}[T_{B}^{2} +
R_{B}^{2} - 2 T_{B}R_{B}\gamma^{2}cos(\phi_{A}-\phi_{B})]\\
\Gamma\{2,3\} \propto R_{B}T_{B}[T_{A}^{2} +
R_{A}^{2} -2 T_{A}R_{A}\gamma^{2}cos(\phi_{A}-
\phi_{B})]\\
\Gamma\{2,4\} \propto [T_{A}^{2}R_{B}^{2}
+T_{B}^{2}R_{A}^{2}+2T_{A}R_{A}T_{B}R_{B}\gamma^{2}cos(\phi_{A}
- \phi_{B})]
\end{align}
\end{subequations}

where, for clarity, we have assumed a perfect
single photon source with $g^{(2)}(0) = 0$ and
perfect spatial overlap of the two input modes at
each coupler. The wave-function overlap of the
two photons is $\gamma = <\psi_{1}|\psi_{2}>$.
From equation [1], we see that if $(\phi_{A} -
\phi_{B}) = 0$ coincidences will only occur for
detectors arranged symmetrically about the
experiment $\{1,3\}$ and $\{2,4\}$ in Fig1(a),
whereas anti-symmetric detector pairs, $\{1,4\}$
and $\{2,3\}$, will not detect coincidences. In
this way, it appears that post-selected events
where Alice and Bob receive one photon each are
entangled in position and time. If $(\phi_{A} -
\phi_{B}) = \pi$ the opposite is true. As the
positions of the two photons are uncertain until
the time at which the measurements are made this
scheme offers a useful way of distributing
entanglement between remote parties. This scheme
is particularly well suited to a demonstration in
optical fibres as it is robust against
decoherence occurring in the long fibre lengths
between coupler $C_{0}$ and the MZIs. A number of
cryptographic schemes have been demonstrated
based on this idea using parametric
down-conversion sources \cite{Tittel}. One
difference we would like to stress is that in
previous works with non-linear crystals both
photons are created simultaneously, with a fixed
phase relationship. Thus interference occurs when
each photon takes the same paths in either
interferometer (either long-long or short-short).
However, in our experiment the two photons are
created at different times and have no fixed
phase relationship. Thus only fourth order
interference occurs when the photons take
opposite paths in the two MZIs.

Note that any variation in $\Delta t$ that is
much less than the photon coherence time does not
affect the results of these correlations.
However, to observe this effect it is necessary
to stabilize $(n(l_{A}-s_{A})/c -
n(l_{B}-s_{B})/c)$ to be fixed for each
measurement and to be able to control this
quantity on a sub-radian scale. Variations in
room temperature, air flow and mechanical stress
will disrupt the MZIs and lead to a loss of
entanglement.

The matching of $\Delta t$ to $n(l_{A}-s_{A})/c$,
and also to $n(l_{B}-s_{B})/c$, could be achieved
using variable delay lines in each MZI but these
components typically induce a sizable photon
loss. Although, wavelength-scale stability can be
achieved with active stabilization and some
feedback system \cite{ZYOptExp} we have opted to
perform the experiment in a modified
interferometer where any shifts in phase occur
equally to both MZIs simultaneously. This
interferometer is shown in Fig. 1(b). In our
interferometer the two photons are once again
split at coupler $C_{0}$ and travel in opposite
directions. However, now the leftward traveling
photon has its polarization changed from H to V.
The two photons are then fed into opposite sides
of a single MZI where they travel through in
opposite directions. As both photons have
orthogonal polarization a polarization-dependent
phase shift can be applied in one arm of the
interferometer, $(\phi_{H} - \phi_{V})$ to
provide the required control of the phase shift
analogous to $(\phi_{A} - \phi_{B})$ in Fig.
1(a). In our experiment any drift then occurs
equally in both the leftward and rightward
traveling photons. The logic behind our labeling
of the couplers $C_{A}$ and $C_{B}$ in Fig. 1(a)
is now clear. If it is assumed that there is a
negligible polarization dependence to $R_{A}$ and
$R_{B}$, we can predict the visibilities of
interference that may be expected between
detector pairs, $V(i,j)$, from equations [1].
Although this is an approximation we have
confirmed experimentally that there is a
significantly larger difference between $R$ and
$T$ values for a given coupler than between the
two $R$ values for different polarizations.

\begin{subequations}
\begin{align}
V(1,3) = \gamma^{2}\\
V(1,4) = 2T_{B}R_{B}\gamma^{2}/(T_{B}^{2} +
R_{B}^{2})\\
V(2,3) = 2T_{A}R_{A}\gamma^{2}/(T_{A}^{2} +
R_{A}^{2})\\
V(2,4) = 2T_{A}R_{A}T_{B}R_{B}\gamma^{2}/(
R_{A}^{2}T_{B}^{2} + T_{A}^{2}R_{B}^{2})
\end{align}
\end{subequations}

Interestingly, due to our symmetric choice of
couplers as defined above we see that $V(1,3)$ is
limited only by the wave-function overlap of the
two photons, $\gamma$.

\begin{figure}
\includegraphics[width=80mm]{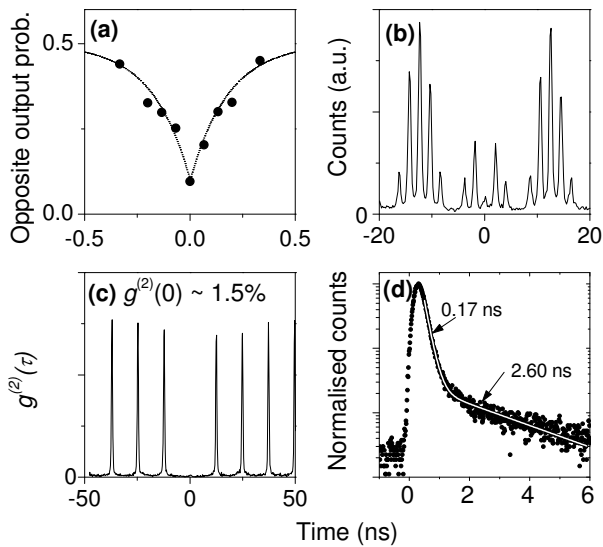}
\caption{\label{fig:epsart} Characteristics of
our single photon source. (a) the Hong-Ou-Mandel
"dip" for this source. The exponential function
is included as a guide for the eye. (b) shows a
correlation recorded between detectors ${3,4}$
using the arrangement in Fig. 1(b), (c) shows the
result of a Hanbury-Brown and Twiss measurement
of the $g^{(2)}(t)$ function for the source and
(d) shows a time resolved trace recorded from the
source under the same excitation conditions.}
\label{Fig2}
\end{figure}

\section{RESULTS}

We now discuss our experimental results. Our
single-photon-source is a pillar microcavity
containing a single quantum dot with exciton
($X$) emission at 943.2nm, coincident with the
cavity $HE_{11}$ mode. The cavity was
photolithographically defined to have a nominal
diameter of 1.75 $\mu$ m, and contained 17 (25)
periods in the Bragg mirror above (below) the
one-wavelength thick cavity. Processing is
carried out with standard photolithography and a
$SiCl_{4}$ reactive ion etch to produce cavities
with $Q$ =4500. During the experiment the sample
is held in a continuous flow cryostat at 3.6K and
optically excited at $\sim 60^{o}$ to the normal
with a ps-pulsed tuneable laser optimised to
excite a resonance in the
photoluminescence-excitation spectrum of the $X$
state, at 908.85nm. The excitation density was
chosen to be more than one order of magnitude
below the level required to saturate emission
from the state so as to minimize dephasing and
temporal jitter associated with occupation of the
biexciton state \cite{BennettOptExp2}. Care is
taken to ensure that the excitation density is
constant throughout the experiment. Photons are
collected with a NA = 0.5 microscope objective on
axis with the pillar. The source delivers photons
with a high multi-photon suppression ratio
($g^{(2)}(0)$ = 1.5\%, Fig. 2(c)) which are close
to the time-bandwidth limit
\cite{BennettOptExp2}. We note that for this
source bunching of photons in adjacent pulses is
negligible. The emission is then spectrally
filtered by a monochromator and the single
photons coupled into our fibre optic
interferometer.

Monitoring correlations between detectors on the
same side of the interferometer, say detectors
$\{1,2\}$, the temporal overlap of the two
photons on $C_{A}$ is optimized by varying the
time delay between the two laser pulses used to
excite the source. A Hong-Ou-Mandel dip is
observed with a visibility of $\sim$ 80\% (Fig,
2(a)).

Using a 4-channel timer card we record
correlations between 4-pairs of detectors
($\{1,3\}, \{1,4\}, \{2,3\}$ and $\{2,4\}$)
simultaneously. For each pair we normalize the
data to the total number of coincidence events
occurring on that channel within $\pm$ 16 ns of
time zero, excluding the central 0.6 ns. This
procedure removes any effect due to varying
singles-count rates on each detector pair (due to
unequal detection efficiencies) and also any
effect due to varying count rates between
measurements at each setting of $(\phi_{H} -
\phi_{V})$ (which occurs due to small changes in
the coupling of photons into the optic fibres).
After this normalization procedure every peak at
a given time delay has the same area in each
correlation. An example is shown in Fig. 4(a) for
channel {1,3} for maximally constructive and
destructive interference. For each channel we
then define $\Gamma\{i,j\}$ as the integrated
area within a 0.6 ns-wide window centered on time
zero for this normalized data. We chose a 0.6ns
time window as this encompasses the central peak
area whilst minimizing the contribution of the
flat background we observe in the correlation
(the origin of which will be discussed later).
All $\Gamma\{i,j\}$ are then normalized by the
same constant, so as to ensure the mean value
$\langle\Sigma_{i,j} \Gamma\{i,j\}\rangle$,
calculated from all measured settings of
$(\phi_{H} - \phi_{V})$, is unity. The results of
this data set are shown in Fig. 3 for each of the
four pairs of detectors. Least squares fits to
the functions stated in equation 1 are shown as
solid lines and from these the visibilities of
the interference, $V_{1}$ are calculated (see
Fig. 3). As expected the highest visibility of
interference is observed on detectors $\{1,3\}$.

\begin{figure}
\includegraphics[width=80mm]{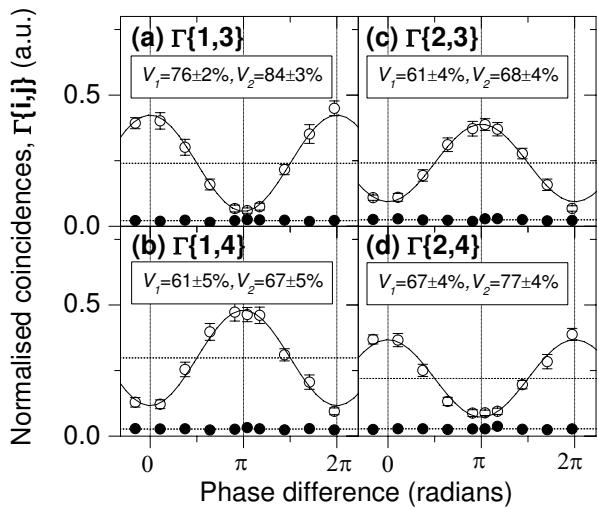}
\caption{\label{fig:epsart} Normalized
coincidences recorded between detector pairs as a
function of the phase setting, $(\phi_{H} -
\phi_{V})$ (open symbols). Estimates of the
background contribution to the correlation counts
as time zero from delayed photon emission events
are also plotted (filled symbols). The labels
show the visibilities of the raw data ($V_{1}$)
and the visibilities after background
subtraction, $V_{2}$. } \label{Fig3}
\end{figure}

We note that because we have unequal reflection
and transmission coefficients in our couplers the
average count rate observed on each channel
differs (shown as a dotted horizontal line in
Fig. 3). On average it is higher for both the
anti-symmetric pairs, and the visibilities are
lower for both of these pairs. From fits to the
measured data we have confirmed that $\sum_{i,j}
\Gamma\{i,j\} = 1$ for each setting of $(\phi_{H}
- \phi_{V})$ within errors. This observation can
only be made because our definition of
$\Gamma\{i,j\}$ is independent of the degree of
correlation on other channels for that particular
setting of $(\phi_{H} - \phi_{V})$. An
alternative method of analysis would have been to
define the sum of the counts in central 0.6ns for
all channels at each setting of $(\phi_{H} -
\phi_{V})$ to be unity.

\section{FACTORS LIMITING THE VISIBILITY}

The experiment was repeated with three separate
quantum dots located on the same semiconductor
chip. In all cases visibilities in excess of 55\%
have been observed, suggesting that more than
half of all post-selected photon pairs are
entangled. For simplicity we have only presented
data from one of these photon sources here but we
have noted that for all these QDs there is a
clear "tail" to the photoluminescence excited
from the sample. Shown in Fig.2(d) is a
time-resolved photoluminescence trace data
recorded from the photon source recorded under
the same excitation conditions. An excellent fit
to this data can be made using two exponential
functions: a fast component which occurs promptly
after the excitation pulse (measured decay time
of 170 ps, limited by our APD time resolution)
and a longer lived component with a decay time of
2.60 ns. Such long-lived components in the $X$
emission from a single QD have been observed
before \cite{Smith, Schwab} and are usually
attributed to the population of dark-states or
charge states in the QD that store carriers for a
time before repopulating the $X$ state. Such
delayed photons are obviously not time-bandwidth
limited and can in fact lead to small number of
false coincidences at time zero.

\begin{figure}
\includegraphics[width=80mm]{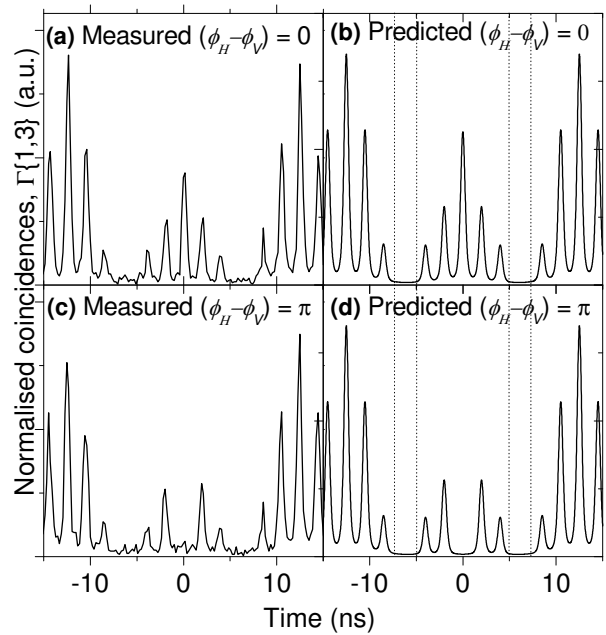}
\caption{\label{fig:epsart} Measured correlation
between $\{1,3\}$ in the case where $(\phi_{H} -
\phi_{V}) = 0$ (a), $(\phi_{H} - \phi_{V}) = \pi$
(c). Also shown are the predicted form of the
correlation for $(\phi_{H} - \phi_{V}) = 0$ (b),
$(\phi_{H} - \phi_{V}) = \pi$ (d) assuming
$g^{(2)}(0) =0$. This illustrates that the long
lived component of the PL emission from the state
contributes a constant background that is $\sim$
the same at time zero as it is in the range 5-7
ns. } \label{Fig4}
\end{figure}

To estimate the number of false counts we might
expect at time zero from this tail we have used
the raw data to reconstruct the correlations we
may expect between a pair of detectors if the
visibility were 100\%. This was achieved by
convolving the experimental time-resolved data in
Fig.2(d) with itself to obtain the shape of the
peaks expected in the autocorrelation. This peak
shape provides an excellent fit to the
peak-shapes in Fig. 2(b) and (c). Multiple peaks,
scaled correctly and offset from time-zero are
then used to construct the expected correlations.
Finally, the effect of dark counts was added into
each plot. The resulting fits are shown in Fig. 4
for both the cases where $(\phi_{H} - \phi_{V}) =
0$ (Fig. 4(b)) and $(\phi_{H} - \phi_{V}) = \pi$
(Fig. 4(d)), which show good agreement with the
corresponding experimental data from (1,3), shown
in Fig. 4(a) and (c). Our purpose in showing the
modeled data is to illustrate that the long lived
component of the emission results in a number of
false counts in the central 0.6 ns window which
is identical to the constant level of false
counts observed between 5.0 and 7.5 ns (and
similar time-windows spaced at the 12.5ns
repetition period). The presence of the
long-lived component to the $X$ state emission
thereby degrades the visibility of the
interference. Therefore, from the raw correlation
data for each channel $\Gamma\{i,j\}$ we have
obtained a direct indication of the contribution
to counts at time zero for every dataset. The
results of this analysis are shown in Fig.3 as
filled-data points on each plot. Subtracting the
effect of these false counts we are able to
obtain a more accurate measure of the visibility
of interference, $V_{2}$, which is stated for
each channel in Fig.3. The visibilities observed
from each detector pair are now either greater
than, or within error equal to, the 71\% value
required prove non-locality \cite{Franson}. As
predicted by equation 2, the correlation observed
between $\{1,3\}$ displays the highest visibility
at 84 $\pm$ 3 \%.

We also expect that the finite value of
$g^{(2)}(0)$ will add to false counts at time
zero. Using the formula quoted for the area of
the central peak by Santori $et$ $al$
\cite{SantoriNature} it is possible to calculate
the overlap of the wavefunctions of the two
photons, $\gamma^{2} = V(1,3).(1 + 2g)$ where $g$
is the probability that the source emits more
than one photon as a result of either one of the
excitation pulses occurring in each 12.5 ns
cycle. In the limit where the excitation level is
well below saturation, as it is here, we assume
that $g \simeq 2g^{(2)}(0)$. Thus we are able to
infer a wavefunction overlap, $\gamma$, of 94
$\pm$ 3\%. The remaining effect reducing this
value from 100 \% is the temporal
distinguishability of the photons emitted from
the source.

\section{CONCLUSION}

To conclude, we have used a single photon source
to generate photons entangled in position and
time. Passive matching of time delays and phase
shifts are achieved by passing the two photons
through the same interferometer in different
directions, with different polarizations. The
degree of wave-function overlap for the two
photons is as high as 94 $\pm$ 3 \% suggesting
that with further improvements these sources may
be of use for quantum information applications,
such as entanglement distribution between remote
stations or as ancilla photon sources in photonic
quantum computing.

This work was partly supported by the EU through
the IST FP6 Integrated Project Qubit Applications
(QAP: contract number 015848) and Network of
Excellence SANDiE.


\end{document}